\documentclass[12pt,a4paper]{article}

\usepackage[english]{babel}
\usepackage{graphicx}
\usepackage[a4paper,left=3cm,right=3cm,top=3cm,bottom=4cm]{geometry}
\usepackage{amsmath,amsfonts,amssymb}
\usepackage{mathrsfs}     
\usepackage{dsfont}      
\usepackage{multirow}
\usepackage{bm}			
\usepackage{cite}
\usepackage{enumerate}
\usepackage{sectsty}

\sectionfont{\large}
\subsectionfont{\normalsize}

\newcommand\Lagr{\mathscr{L}}
\newcommand\identity{\mathds{1}}
\renewcommand\d{\textrm{d}}
\renewcommand{\bar}{\overline}
\DeclareMathOperator{\tr}{Tr}
\newcommand\T{\textrm{T}}

\begin{document}

\title{Dangerous Skyrmions in Little Higgs Models}
\author{Marc Gillioz}

\thispagestyle{empty}

\begin{flushright}
ZU-TH 21/11 \\
LPN11-57
\end{flushright}

\vspace{3cm}

\begin{center}

{\Large {\bf Dangerous Skyrmions in Little Higgs Models}}

\vspace{2cm}

{\large Marc Gillioz}

\vspace{1cm}

{\it
Institute for Theoretical Physics, University of Zurich, \\
Winterthurerstrasse 190, CH-8057 Z\"urich, Switzerland
}

\end{center}

\vspace{2cm}

\begin{abstract}

Skyrmions are present in many models of electroweak symmetry breaking where the Higgs is a pseudo-Goldstone boson of some strongly interacting sector. They are stable, composite objects whose mass lies in the range 10--100~TeV and can be naturally abundant in the universe due to their small annihilation cross-section. They represent therefore good dark matter candidates. We show however in this work that the lightest skyrmion states are electrically charged in most of the popular little Higgs models, and hence should have been directly or indirectly observed in nature already. The charge of the skyrmion under the electroweak gauge group is computed in a model-independent way and is related to the presence of anomalies in the underlying theory via the Wess-Zumino-Witten term.

\end{abstract}

\vfill

\def\thefootnote{\arabic{footnote}}
\setcounter{page}{0}
\setcounter{footnote}{0}

\newpage


\section{Introduction}

The composite Higgs idea~\cite{Kaplan:1983fs} is an interesting alternative to supersymmetry regarding the hierarchy problem of the Standard Model, in which the Higgs boson is a pseudo-Goldstone boson resulting from the spontaneous symmetry breakdown of some new sector. It is the mechanism behind the numerous little Higgs models~\cite{ArkaniHamed:2001nc, ArkaniHamed:2002qx, ArkaniHamed:2002qy, Low:2002ws, Chang:2003un, Kaplan:2003uc, Cheng:2003ju, Cheng:2004yc, Schmaltz:2010ac} and other recent composite models arising from a holographic description~\cite{Agashe:2004rs, Gripaios:2009pe}. In all these models, new particle states live at a scale $f$ above the experimental bounds, while the approximate global symmetry of this new sector is spontaneously broken to a subgroup of it, yielding Goldstone bosons naturally lighter than $f$, among which a complex doublet can be identified with the Higgs. The phenomenology of the new particle states living at the TeV scale and below has been extensively studied in the last few years (see for example~\cite{Giudice:2007fh}). Heavier states might nevertheless play an equally important role, as long as they are stable over cosmological timescale. This is the case of the skyrmions which we are studying here.

The Higgs sector of composite models is described at energies below the cutoff $\Lambda \sim 4 \pi f$ by a non-linear sigma-model, with a scalar field $\Phi$ parametrising the Higgs and possibly other light scalars. While the generation of quark and lepton masses is very model-dependent, the couplings of the Higgs fields to its new partners and electroweak gauge bosons are completely determined by the symmetry breaking pattern and the embedding of gauge fields in the global symmetry group. The description in terms of a non-linear sigma-model mimics the low-energy chiral effective theory of QCD, in which the pseudo-Goldstone states are the pions. Some of theses sigma-models contain topological solitons, depending on the homotopy of the vacuum manifold. In little Higgs models, apart from cosmic strings and monopoles~\cite{Trodden:2004ea}, skyrmions~\cite{Skyrme:1961vq, Witten:1983tx} might also be present. In the case where the underlying theory is a strongly-coupled theory of (techni-)fermions\footnote{For convenience, we will always refer to the fermions of the new strongly-interacting sector as \emph{technifermions}, even if the underlying theory does not necessarily share many features with technicolor. Similarly, the term \emph{composite Higgs} refers to any realisation of the Higgs as a pseudo-Goldstone boson, including little Higgs models.} --- as for QCD --- these skyrmions can be identified with the (techni-)baryons of the theory. The physical meaning of skyrmions in models with a weakly coupled UV completion is less understood.

In this work we study the properties of skyrmions in a general framework describing the composite Higgs models, and in particular their charge under the electroweak gauge group. Since skyrmions arise as non-perturbative objects, it is in general not possible to compute their mass and size in terms of the parameters of the low-energy effective theory only; a complete knowledge of the UV completion of the model would be needed. The quantum properties of skyrmions seem also to be dependent on the underlying theory, since they appear as bound states of technifermions. We argue nevertheless in the following that the charge of the skyrmion under the electroweak gauge group is very much constrained by the symmetry breaking pattern and the presence of gauge fields, and can be determined without knowing the UV completion of the model.

The presence of skyrmions as very massive and stable objects in composite Higgs models has raised a lot of interest since they naturally provide a dark matter candidate. This idea was already discussed long ago about the technibaryons of technicolor~\cite{Nussinov:1985xr}, and recent studies have also shown that skyrmion can play the role of dark matter in composite Higgs models in general~\cite{Murayama:2009nj}, and more particularly in little Higgs models~\cite{Joseph:2009bq, Gillioz:2010mr}. The skyrmion appears as an new stable particle as an alternative --- or in addition --- to the one resulting from the presence of a discrete $\mathbb Z_2$ symmetry such as T-parity in little Higgs models~\cite{Cheng:2003ju}. The possibility of skyrmions playing the role of dark matter is however only plausible if the lightest skyrmion state is electrically neutral. The very point of this work is to show that this cannot always be achieved in little Higgs models. We demonstrate that the lightest skyrmion state is always formed as a doublet of $SU(2)_W$, and that its hypercharge is fixed by the embedding of the $U(1)_Y$ generator in the global symmetry group of the theory. The hypercharge of the skyrmion is sensitive to the anomalies of the underlying theory, expressed in the low-energy description by the Wess-Zumino-Witten term~\cite{Wess:1971yu,Witten:1983tw}. This very term is already famous for breaking T-parity in little Higgs models~\cite{Hill:2007nz, Hill:2007zv}, although a different implementation of a discrete parity might still be realised~\cite{Krohn:2008ye, Freitas:2009jq}. Our results point therefore to a new issue in little Higgs models in addition to existing problems~\cite{Schmaltz:2008vd, Pappadopulo:2010jx}, and accentuate the difficulty of constructing a fully realistic model.

This work is structured as follows. In section~2 we discuss the general form of the Higgs sector in composite Higgs models and study the skyrmion in a minimal setup containing all the relevant features. This includes constructing the classical skyrmion solution, quantising it, and deriving the quantum numbers of the lightest skyrmion states. We also show how to extend the minimal model in order to obtain electrically neutral skyrmions. In section~3 we review the other possible symmetry breaking patterns and discuss the electric charge of the skyrmion, when one is present. Special care is given to the littlest Higgs model~\cite{ArkaniHamed:2002qy}, which exhibits quite special features in comparison to the other cases. Finally we summarise our results and conclude by discussing the consequences of having charged skyrmions in little Higgs models.


\section{Skyrmions in the presence of the electroweak gauge group}
\label{sec:toymodel}

We consider a theory containing a new, strongly-interacting sector in which a global symmetry $G$ is broken down to a subgroup $H$ of it, yielding a certain number of Goldstone bosons, among which the Higgs doublet. At energies below the symmetry breaking scale $f$, the Higgs sector is described by a non-linear sigma-model in the form
\begin{equation}
	\Lagr = \frac{f^2}{4} \tr\left( D_\mu \Phi D^\mu \Phi^\dag \right) + \textrm{higher order terms},
	\label{eq:sigmamodel}
\end{equation}
where $\Phi(x)$ is the Goldstone field and $f$ parametrises the scale of new physics, typically of the order of a few TeV. This description of the Higgs field is common to both little Higgs and holographic Higgs models. The global symmetry group $G$ of the strong sector is furthermore explicitly broken by gauging at least a $SU(2)_W \times U(1)_Y$ subgroup of it; realistic models often require to gauge a larger subgroup. The gauged electroweak subgroup of $G$ has to be chosen carefully so that the Goldstone field $\Phi$ contains an $SU(2)_W$ doublet with hypercharge $\frac{1}{2}$ to be identified with the Higgs boson. The gauge bosons couple to the Goldstone sector through the covariant derivative. In the absence of electroweak couplings $g_2$, $g_1$ and of Yukawa couplings to the Standard Model fermions, the Higgs would be massless, as an exact Goldstone boson. Only radiative corrections trigger the electroweak symmetry breaking and give a mass to the $W$ and $Z$ bosons as well as to the Higgs.

\subsection{A minimal model}

For simplicity, we will first restrict our discussion to a minimal model, then show that the lessons we learn from it can be extended to more general cases. The simplest realisation of the Higgs as a pseudo-Goldstone boson consists in taking $\Phi \in SU(3)$. This can be achieved by adding a new strongly interacting sector to the Standard Model, composed of 3 families of fermions, each of them coming in $N_c$ colors. The Standard Model gauge group is therefore enlarged with a new $SU(N_c)$, under which the new fermions transform in the fundamental representation. In the absence of fermion masses, there is a global $SU(3)_L \times SU(3)_R$ symmetry rotating the left- and right-handed fermions of the 3 families. This symmetry is spontaneously broken down to the diagonal subgroup $SU(3)_V$ by fermion condensates. Note that realistic models based on the symmetry breaking pattern $SU(3)_L \times SU(3)_R \to SU(3)_V$ actually require the introduction of several copies of the global symmetry group, as in the ``minimal moose'' model~\cite{ArkaniHamed:2002qx}.

To correctly describe the anomalies induced by loops of fermions of the strongly interacting sector, the Wess-Zumino-Witten (WZW) term~\cite{Wess:1971yu,Witten:1983tw} has to be added to the Lagrangian~(\ref{eq:sigmamodel}). This term cannot be expressed in terms of local operators, and thus has to be defined on a five-dimensional manifold $\mathcal{M}_5$ whose boundary is the usual four-dimensional Minkowski space: 
\begin{eqnarray}
	\Gamma_{WZW} & = & -\frac{i N_c}{240 \pi^2} \int_{\mathcal{M}_5} \d^5x \, \epsilon^{\mu\nu\rho\sigma\tau}
		\tr\left( \Phi^\dag \partial_\mu \Phi \, \partial_\nu \Phi^\dag \partial_\rho \Phi \,
			\partial_\sigma \Phi^\dag \partial_\tau \Phi \right) \nonumber \\
	&& + \textrm{~local terms required by gauge invariance,}
	\label{eq:WZWterm}
\end{eqnarray}
where $N_c$ is the number of colors of the underlying strongly-interacting theory. The non-local five-dimensional part of the WZW term is not gauge invariant, but its variation under gauge transformations is local and can be compensated by a four-dimensional term written in terms of gauge fields and of the Goldstone field $\Phi$~\cite{Kaymakcalan:1983qq}. The exact form of this term depends on the choice of the gauge group; it is illustrated in the appendix for a vector gauge group. The number of colors $N_c$ can in principle be taken to zero, in which case the underlying theory is not anymore a theory of strongly-interacting fermions and the scalars described by the sigma-model~(\ref{eq:sigmamodel}) are fundamental scalars. Note, however, that such a weakly-coupled UV completion might reintroduce the hierarchy problem.

In our minimal model the field $\Phi(x)$ transforms in the adjoint representation of $SU(3)_V$. The covariant derivative used in the Lagrangian~(\ref{eq:sigmamodel}) is given by
\begin{equation}
	D_\mu \Phi = \partial_\mu \Phi - i g W_\mu^a \left[ Q^a, \Phi \right] - i g' B_\mu \left[ Y, \Phi \right].
	\label{eq:covariantderivative}
\end{equation}
The commutators ensure that the electroweak gauge group belongs to the unbroken diagonal subgroup $SU(3)_V$. This description in terms of the low-energy effective theory does not fix the charge of the technifermions under the electroweak gauge group: the diagonal $SU(2)_W \times U(1)$ subgroup of the global $SU(3)_L \times SU(3)_R$ symmetry can be gauged as such, but can also arise as the unbroken combination of a larger gauge group, for example in the form $SU(2)_L \times SU(2)_R \times U(1)_L \times U(1)_R$. A remarkable result of our analysis is indeed that the skyrmion charge does not directly depend on the technifermion content of the theory, as explained in the following. The generators of the electroweak $SU(2)_W$ gauge subgroup can be taken without loss of generality to live in the upper-left $2 \times 2$ block of the $3 \times 3$ matrix, as
\begin{equation}
	Q^a = \frac{1}{2} \left( \begin{array}{cc}
		\sigma^a & \\
		& ~ 
	\end{array} \right),
	\label{eq:SU2gaugegenerators}
\end{equation}
where the $\sigma_a$ are the usual Pauli matrices. The hypercharge generator has to commute with the $Q^a$ and be traceless, it is therefore fixed up to an overall factor to 
\begin{equation}
	Y = \frac{1}{6} \left( \begin{array}{cc}
		\identity & \\
		& -2
	\end{array} \right).
	\label{eq:U1gaugegenerator}
\end{equation}
The $\frac{1}{6}$ factor is chosen to obtain among the Goldstone bosons a $SU(2)_W$ doublet with the quantum numbers of the Higgs.\footnote{Note that unlike for the $SU(2)_W$ generators $Q_a$, we do not require the hypercharge generator $Y$ to satisfy the normalisation condition $\tr Y Y = \frac{1}{2}$. This permits to write the covariant derivative in the form of eq.~(\ref{eq:covariantderivative}), i.e.~without an extra factor in front of the term $B_\mu \left[ Y, \Phi \right]$. In other words, the hypercharge of the Goldstone field $\Phi$ is directly encoded into the generator $Y$.} The Goldstone field $\Pi$, given in the nonlinear realisation of the sigma-model by
\begin{equation}
	\Phi = \exp\left[ 2 i \Pi / f \right],
\end{equation}
can then be decomposed as
\begin{equation}
	\Pi = \left( \begin{array}{cc}
		\boldsymbol\omega + \frac{1}{2\sqrt{3}} \, \eta  & h \\
		h^\dag & -\frac{1}{\sqrt{3}} \, \eta
	\end{array} \right),
	\label{eq:Goldstonematrix}
\end{equation}
where $\boldsymbol\omega$ is a real triplet, $h$ a complex doublet and $\eta$ a real singlet of $SU(2)_W$, denoted by
\begin{equation}
	\boldsymbol\omega = \left( \begin{array}{cc}
		\frac{1}{2} \, \omega^0 & \frac{1}{\sqrt{2}} \, \omega^+ \\
		\frac{1}{\sqrt{2}} \, \omega^-  & -\frac{1}{2} \, \omega^0
	\end{array} \right),
	\hspace{1cm}
	h = \frac{1}{\sqrt{2}} \left( \begin{array}{c}
		h^+ \\ h^0
	\end{array} \right),
	\hspace{1cm}
	\eta = \eta^0.
	\label{eq:Higgsdoublet}
\end{equation}
Note that $\boldsymbol\omega$ and $\eta$ have automatically zero hypercharge, independently of the choice of the normalisation of the generator $Y$, while this is not the case for $h$. The generator of the electric charge is then given by
\begin{equation}
	Q_{em} = Q^3 + Y = \left( \begin{array}{ccc}
		\frac{2}{3} && \\ & -\frac{1}{3} & \\ && -\frac{1}{3}
	\end{array} \right),
	\label{eq:electricchargegenerator}
\end{equation}
and turns out to be the same as in the low-energy chiral description of QCD, with the $u$, $d$ and $s$ quarks taken as massless. For this reason, this model is also a plausible composite Higgs candidate, since the charges of the new technifermions match the ones of the Standard Model quarks, and one could choose not to couple the latter directly to the Higgs sector, but still obtain a mass by mixing with the heavy fermions resonances.\footnote{This would however require the strong sector to be the usual color $SU(3)_c$, hence fixing $N_c = 3$ in the WZW term~(\ref{eq:WZWterm}).}

\subsection{Topological charge}

The presence of skyrmions in this model is due to the fact that the third homotopy group of the coset space $SU(N)$ is non-trivial. More precisely $\pi_3(SU(N)) = \mathbb{Z}$, and each field configuration is parametrised by an integer winding number, or topological index. For $SU(N)$-valued fields, this topological index can moreover be expressed as an integral over space as
\begin{equation}
	\eta(\Phi) = \frac{1}{24 \pi^2} \epsilon_{ijk} \int \d^3x \tr \left( \Phi^\dag \partial_i \Phi \, \partial_j \Phi^\dag \partial_k \Phi \right)
		\in \mathbb{Z}.
	\label{eq:windingnumber}
\end{equation}
The integrand in the previous equation is not obviously invariant under gauge transformations. However, let's consider the gauge invariant current
\begin{equation}
	B^\mu = \frac{1}{24 \pi^2} \epsilon^{\mu\nu\rho\sigma} \left[ \tr \left( \Phi^\dag D_\nu \Phi \, D_\rho \Phi^\dag D_\sigma \Phi \right)
		+ 3 i g \tr F_{\nu\rho} \left( \Phi \, D_\sigma \Phi^\dag - \Phi^\dag D_\sigma \Phi \right) \right]
	\label{eq:topologicalcurrent}
\end{equation}
where $F_{\mu\nu}$ and $g$ denote generically the field strength tensor and coupling of both $SU(2)_W$ and $U(1)_Y$. The conserved charge associated with this current can be written as
\begin{equation}
	B = \int \d^3x ~ B^0 = \eta(\Phi) + \int \d^3x ~\partial_i \Omega_i,
	\label{eq:topologicalcharge}
\end{equation}
where
\begin{equation}
	\Omega_i = \frac{1}{8 \pi^2} \epsilon_{ijk} \left[ i g \tr A_j \left( \Phi \, \partial_k \Phi^\dag - \Phi^\dag \partial_k \Phi \right)
		+ g^2 \tr \left( A_j \, \Phi \, A_k \, \Phi^\dag \right) \right].
\end{equation}
The second term on the right-hand side of~(\ref{eq:topologicalcharge}) is a surface term, and vanishes provided that the fields are in the vacuum at the spatial boundaries. The gauge-invariant charge $B$ is therefore equivalent to the topological index $\eta(\Phi)$, and is conserved in time.\footnote{The identification $\eta = B$ is only valid as long as no singularities are present; topological defects such as monopoles can induce skyrmion decay~\cite{Callan:1983nx}.} In other words, the presence of gauge fields in a vector representation do not modify the topological structure of the model. On the contrary, gauge fields associated with a symmetry which is broken by the vacuum expectation value $\langle\Phi\rangle$ can lead to skyrmion decay, via an instanton~\cite{D'Hoker:1983kr}. The tunnelling probability of such a process is nevertheless exponentially suppressed, so that the skyrmions remain stable anyway on cosmological timescales.

The different sectors of the space of field configurations are labelled by the integer-valued winding number~(\ref{eq:windingnumber}). The vacuum configuration $\Phi = \identity$ and any continuous transformation of it correspond to the topologically trivial sector of the theory with $B = 0$. Field configuration of non-zero winding number can be obtained from a \emph{hedgehog ansatz},
\begin{equation}
	\Phi_h = \exp\left[ 2 i \, F(r) \, \hat x_i \, T^i \right],
	\label{eq:hedgehog}
\end{equation}
where $F(r)$ is a function of the radial variable $r = \sqrt{x_i^2}$, the $\hat x_i = x_i / r$ are angular coordinates, and the $T^i$ are generators forming a $\mathfrak{su}(2)$ algebra and satisfying a proper normalisation condition,
\begin{equation}
	\left[ T^a, T^b \right] = i \, \epsilon^{abc} \, T^c,
	\hspace{1cm}
	\tr\left( T^a T^b \right) = \frac{1}{2} \delta^{ab}.
	\label{eq:su2algebra}
\end{equation}
The winding number for the hedgehog field configuration depends then on the boundary conditions for the function $F$ only. Choosing $F(0) = k \pi$, $F(\infty) = 0$ yields $B(\Phi_h) = k$. Note also that the hedgehog ansatz~(\ref{eq:hedgehog}) is spherically symmetric, in the sense that a $SO(3)$ rotation in space acting on the angular coordinates $\hat x_i$ is equivalent to a $SU(2)$ transformation acting on the generators $T^i$, as a subgroup of $SU(3)_V$. As a consequence, since global $SU(3)_V$ transformations are symmetries of the model, the energy density obtained with the hedgehog ansatz is explicitly independent of the angular coordinates: the skyrmion has a spherical form.

\subsection{Stability, Skyrme term and Bogomolny bound}

The non-trivial topology of the vacuum manifold is actually not enough to ensure the presence of skyrmions. In the absence of higher-order terms in the Lagrangian~(\ref{eq:sigmamodel}), the size of the topological soliton would actually shrink to zero, following Derrick's theorem~\cite{Derrick:1964ww}. The presence of at least one higher-derivative operator is sufficient to guarantee skyrmions field configurations of finite size and mass at the classical level.\footnote{Since the skyrmion is a static solution of the field equation, each term in the Lagrangian contributes positively to its energy. The fundamental requirement for the skyrmion stability is therefore that there are at least two operators in the Lagrangian with different mass dimensions, so that the balance between their scaling can define a preferred scale for the skyrmion size.} In general, we can assume that these higher-order term are present in our models, but one cannot know their precise form from the point of view of the low-energy theory. The mass of the skyrmion may depend dramatically on the coefficients of higher-order term and cannot be computed without fixing them.

It is important to note here that there exists models in which none of these higher-order terms is present, and for which no stable skyrmions can exist. This scenario is realised in some little Higgs models where the scalar field playing the role of the Higgs is an elementary field, described before spontaneous symmetry breaking by a linear sigma-model~\cite{Batra:2004ah, Csaki:2008se}. In this case, another mechanism is then required at a higher scale to control the mass of the fundamental scalar, which may be supersymmetry.

For the problem adressed in this work, the relevant properties of the skyrmion are its symmetries (both global and local), but not its mass, size, or other static properties. We will therefore get around the problem of higher-order operators by first discussing a simpler model where only the Skyrme term~\cite{Skyrme:1961vq} is present, and then show that our conclusions are actually independent of the exact form of the sigma-model Lagrangian. The Skyrme term contains four derivatives, and can be expressed with the help of commutators as
\begin{equation}
	\Lagr_\textrm{Skyrme} = \frac{1}{32 e^2} \tr \left[ \Phi^\dag D_\mu \Phi, \Phi^\dag D_\nu \Phi \right]
		\left[ \Phi^\dag D^\mu \Phi, \Phi^\dag D^\nu \Phi \right].
	\label{eq:Skyrmeterm}
\end{equation}
Taking just this term in the derivative expansion has many advantages. First, due to the antisymmetrisation of the Lorentz indices, the Skyrme term contains at most two time derivatives, which is a necessary condition to perform a canonical quantisation of the skyrmion.\footnote{The Skyrme term is not the only term with two time derivatives. In the literature, the square of the topological current~(\ref{eq:topologicalcurrent}) is often used as well~\cite{Adkins:1983nw, Jackson:1985yz}.} Second, all skyrmion properties will depend only on the two unknown parameters $f$ and $e$.\footnote{The other two parameters of the model, the gauge coupling $g$ and $g'$, are here implicitly fixed to the Standard Model values.} Third, the energy of every field configuration within this model is bounded from below by the quantity~\cite{Brihaye:1998,Brihaye:2004pz}:
\begin{equation}
	E[\Phi] \geq 6 \pi^2 \frac{f}{e} \left| B \right| \left[ 1 + \left( \frac{3 g}{2 e} \right)^2 \right]^{-\frac{1}{2}},
	\label{eq:Bogomolnybound}
\end{equation}
where $B$ is the topological charge. In the limit $g \to 0$, the usual Bogomolny bound for an ungauged skyrmion is recovered. This bound cannot be saturated but gives already a good estimation of the skyrmion mass. It shows also that the presence of gauge fields allows to lower the mass of the skyrmion, as we will see in the next section.

\subsection{The Skyrme-Wu-Yang ansatz}

In the absence of gauge fields, the hedgehog ansatz~(\ref{eq:hedgehog}) is known to yield the lowest energy configuration of unit winding number in the Skyrme model~\cite{Adkins:1983ya}. Any embedding of the hedgehog into the $SU(3)$ field $\Phi$ is equivalent, i.e.~any choice of $\mathfrak{su}(2)$ generators $T^a$ is allowed. Turning on the $SU(2)_W \times U(1)_Y$ electroweak gauge group removes this degeneracy. The crucial point is now to find the form of the lightest field configuration of unit winding number. Non-vanishing gauge fields give a positive contribution to the energy of a skyrmion configuration through their kinetic term, but they can still lower the overall energy by their interplay with the scalar field $\Phi(x)$.

The effect of a $U(1)$ gauge field on the skyrmion mass has been studied in the literature~\cite{Piette:1997ny}. A solution with non-vanishing gauge field is favoured, independently of the strength of the gauge coupling. The field configuration in this case is not spherically symmetric anymore, but still preserves an axial symmetry.\footnote{Such a field configuration can be achieved in our model by choosing the $T^a$ to be the generators of a $SU(2)$ subgroup living in the upper-left block of the $SU(3)$ matrix $\Phi$ and considering the gauge field of electromagnetism only. The resulting decrease in the energy of the classical skyrmion configuration with respect to the ungauged hedgehog is shown on figure~\ref{fig:M0} as a function of the Skyrme parameter $e$.} In general, any of the three components of the $SU(2)_W$ gauge field can play the same role and lower the mass of the skyrmion while spoiling its spherical symmetry. The lightest field configuration is however obtained using all three components of $W_\mu^a$. One can take the hedgehog ansatz~(\ref{eq:hedgehog}) to live along the $SU(2)_W$ gauged subgroup, i.e. choose the hedgehog generators to match the gauge ones
\begin{equation}
	T^a = Q^a.
	\label{eq:generatorsalignement}
\end{equation}
With this choice, the spherical symmetry of the skyrmion can be preserved, since $SU(2)_W$ gauge transformations acts then in a similar fashion as spatial rotations on the hedgehog configuration. The so-constructed hedgehog lives exclusively in the upper-left $2 \times 2$ block of the matrix $\Phi(x)$, and commutes with the hypercharge generator. Thus the $U(1)_Y$ gauge field cannot help lowering the energy and is set to zero, $B_\mu = 0$. An ansatz for the $SU(2)_W$ gauge fields can be made in the form
\begin{equation}
	W_i^a = \frac{a(r)}{2 g r} \epsilon_{iak} \, \hat x_k,
	\hspace{1cm}
	W_0^a = 0,
	\label{eq:Skyrme-Wu-Yang}
\end{equation}
where $a(r)$ is a function playing the same role as $F(r)$. This is the so-called \emph{Skyrme-Wu-Yang ansatz}~\cite{Brihaye:1998,Brihaye:2004pz}. The boundary conditions for $a(r)$ have to be fixed to $a(0) = 0$ for definiteness at the origin, while at spatial infinity $a(r)$ is only required to take a constant value, i.e.~$a'(\infty) = 0$.

\begin{figure}
	\centering
	\includegraphics[width=0.7\linewidth]{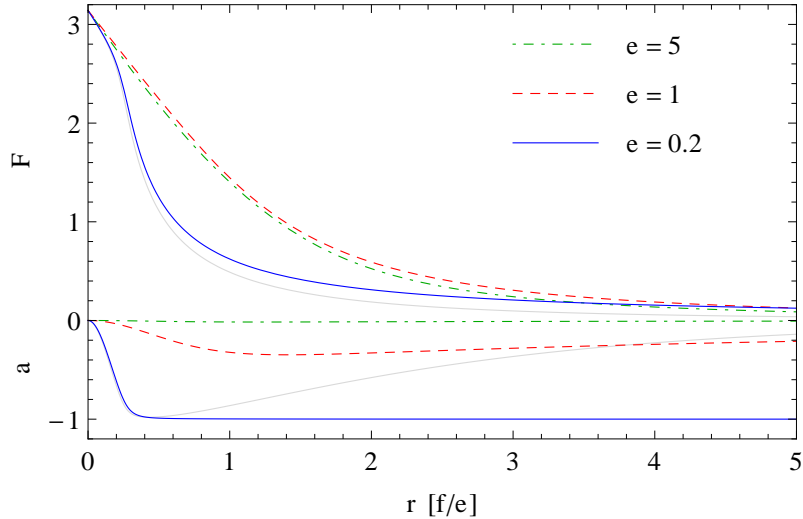}
	\caption{The functions $F$ and $a$ minimising the energy functional~(\ref{eq:energyfunctional}) for different values of the Skyrme coupling $e$, while $g$ is taken to be the Standard Model value $g=0.653$. The grey solid line corresponds to $e = 0.2$ when the electroweak symmetry breaking term~(\ref{eq:epsilonterm}) is turned on (with $\epsilon = 0.1$).}
	\label{fig:F-a}
\end{figure}

Using the Skyrme-Wu-Yang ansatz, the energy of the skyrmion configuration can be written as a functional of $F(r)$ and $a(r)$ as
{\setlength\arraycolsep{1.4pt}
\begin{eqnarray}
	E[F,a] & = & 2 \pi \frac{f}{e} \int\limits_0^\infty \d r
		\left[ \frac{e^2}{g^2} \left( 2 \left( a' \right)^2 + \frac{a^2 (a+2)^2}{r^2} \right)
			+ \left( r^2 + 2 (1+a)^2 \sin^2 F \right) \left( F' \right)^2 \right. \nonumber \\
	&& \hspace{4cm} \left. + (1+a)^2 \sin^2 F \left( 2 + (1+a)^2 \frac{\sin^2 F}{r^2} \right) \right].
	\label{eq:energyfunctional}
\end{eqnarray}}%
The functions $F$ and $a$ minimising this integral for different values of the Skyrme coupling $e$ are shown on figure~\ref{fig:F-a}. In the limit of vanishing gauge coupling, or equivalently for a large $e / g$ ratio, $F$ tends towards the ungauged solution, while $a$ remains small. On the contrary, for values of $e / g \lesssim 1$, the gauge potential decreases as $\frac{1}{r}$, as the function $a$ goes asymptotically to the constant value $-1$. In the latter case the $SU(2)_W$ gauge field mimics at large distances the field induced by a magnetic monopole. The energy of the skyrmion configuration is shown on figure~\ref{fig:M0} (corresponding to $\epsilon = 0$).

The property of the Skyrme-Wu-Yang ansatz to yield always the classical field configuration of lowest energy is expected to be valid independently of the exact form of the higher order terms in the sigma-model, and therefore to be universal.\footnote{In principle a $U(1)$-gauged skyrmion might be lighter than a $SU(2)$-gauged one, provided that the coupling constant associated with the $U(1)$ gauge field is much larger that the one associated with $SU(2)$. This situation does not happen when considering the electroweak gauge group of the Standard Model.} It is another example of the fact that the lowest energy soliton configurations are generally the ones with the highest symmetry properties. The spherical symmetry of the Skyrme-Wu-Yang ansatz is naturally preserved by all terms in the derivative expansion due to the gauge invariance of the model, as long as the alignment of the gauge and hedgehog generators is chosen, as in eq.~(\ref{eq:generatorsalignement}).

The two limiting cases $e \to 0$ and $e \to \infty$ are somewhat special. In the former limit, the gauge fields play a very important role, up to the point where the lightest energy solution takes the asymptotic form of a magnetic monopole; the transition to this regime is visible on figure~\ref{fig:M0} around $e \approx 0.75$ where the energy density shows a knee. Conversely, in the limit of large $e$ (vanishing Skyrme term), the gauge field becomes very weak, and the various skyrmion configurations are nearly degenerate in energy. We want to emphasise however that both limits seem unphysical once all terms in the derivative expansion are taken into account. The optimal value of $e$ in the original chiral Lagrangian of QCD is indeed precisely in the intermediate range, namely $e \approx 4$, in order to predict the correct baryon properties.

\subsection{Electroweak symmetry breaking}

\begin{figure}
	\centering
	\includegraphics[width=0.7\linewidth]{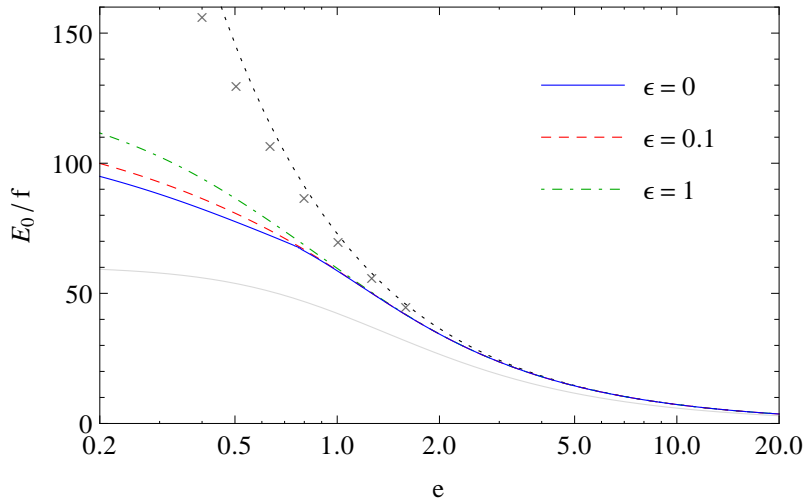}
	\caption{Energy of the skyrmion configuration given by equation~(\ref{eq:energyfunctional}) as a function of the Skyrme parameter $e$, for different values of $\epsilon = v^2 / f^2$. The dotted black line corresponds to the ungauged case, the solid grey one to the Bogomolny bound given in eq.~(\ref{eq:Bogomolnybound}). The crosses show for comparison the mass of the $U(1)_{em}$ gauged solution as computed in ref.~\cite{Piette:1997ny}.}
	\label{fig:M0}
\end{figure}

The discussion so far was based onto the assumption that the vacuum expectation value of the field $\Phi(x)$ is strictly equal to the identity matrix, $\langle\Phi\rangle = \identity$. In this case however the Goldstone bosons would be exactly massless and the electroweak symmetry group would remain unbroken. In realistic models a mass for the Higgs boson is actually obtained by radiative corrections. The details of this mechanism depend on the exact realisation of the composite Higgs model, and we do not want to discuss them here. In our formalism, the breaking of the electroweak symmetry manifest itself as a vacuum misalignment conveniently expressed in terms of the Goldstone field~(\ref{eq:Goldstonematrix}) as
\begin{equation}
	\langle \Pi \rangle = \left( \begin{array}{ccc}
		~ && \\ && v \\ & v &
	\end{array} \right),
\end{equation}
where $v \approx 246$~GeV is the electroweak symmetry breaking scale. This vacuum expectation value yields through the kinetic term for the field $\Phi$ a mass term for the $W$ and $Z$ bosons, of the form
\begin{equation}
	m_W = \sin\left( \frac{v}{2 f} \right) g f \approx \frac{g v}{2},
	\hspace{0.8cm}
	m_Z = \frac{1}{2} \sin\left( \frac{v}{f} \right) \frac{g f}{\cos\theta_W} \approx \frac{g v}{2 \cos\theta_W},
\end{equation}
where $\theta_W$ denotes the Weinberg angle. In the right-hand side equalities we have taken the limit $v \ll f$, recovering the Standard Model values. It is evident that the breaking of the electroweak symmetry also automatically breaks the spherical symmetry of the Skyrme-Wu-Yang ansatz. The correction to the energy of the skyrmion induced by the vacuum expectation value $v$ can be expressed as a correction to the energy density~(\ref{eq:energyfunctional}) in the form
\begin{equation}
	E[F,a] = 2 \pi \frac{f}{e} \int\limits_0^\infty \d r \left[ \ldots + \frac{\epsilon}{2} \, a^2
		+ \mathcal{O}\left( \epsilon^2 \right) \right],
	\label{eq:epsilonterm}
\end{equation}
where $\epsilon = v^2 / f^2$, and the dots denote the $\epsilon$-independent terms of eq.~(\ref{eq:energyfunctional}). Note that the leading order contribution in $\epsilon$ preserves the spherical symmetry of the skyrmion; higher order terms do not. The true gauge group of the Standard Model being $U(1)_{em}$, no spherically symmetric skyrmion configuration including gauge fields can actually be constructed. However, the very small size of the correction introduced in eq.~(\ref{eq:epsilonterm}) implies that the lowest energy configuration is still provided by the Skyrme-Wu-Yang ansatz. The contribution to the energy density coming from the zeroth order in $v/f$ are of order 10--100~$f/e$. Leading order corrections in $v/f$ have already very little effect, since the correction term is bounded by $|a(r)| \leq 1$; the difference is only significant in the regime of small $e$, but remains small even in the unrealistic limit $\epsilon \to 1$, as shown in figure~\ref{fig:M0}. The next-to-leading order terms breaking the spherical symmetry have therefore a negligible effect on the mass of the skyrmion. The $\epsilon$-term plays nevertheless a important role in the sense that it forbids monopole-like solutions, for which the asymptotic value of the function $a(r)$ is different from zero at spatial infinity; its effects on the profile functions $F$ and $a$ are visible on figure~\ref{fig:F-a} for $e = 0.2$ and $\epsilon = 0.1$.

\subsection{Quantisation of the gauged skyrmion}

The physical skyrmion states are obtained, following Witten et al.~\cite{Adkins:1983ya}, by performing a zero-mode quantisation around the classical solution. The zero-modes are transformations that leave invariant the energy of the skyrmion solution found in the previous section. This comprises the Lorentz transformations and the internal symmetries of the skyrmion. The latter reduce after electroweak symmetry breaking to the $U(1)_{em}$ global transformations. However, the correction to the spherical symmetric skyrmion are in our case so negligible compared to the overall skyrmion energy that all $SU(2)_W$ and $U(1)_Y$ global transformations are approximate symmetries of the classical skyrmion, and will be considered as exact in the following. The quantisation procedure consists in promoting the parameters of such transformations to time-dependent variables, compute the Hamiltonian of the system, and finally perform a canonical quantisation.

\begin{figure}
	\centering
	\includegraphics[width=0.7\linewidth]{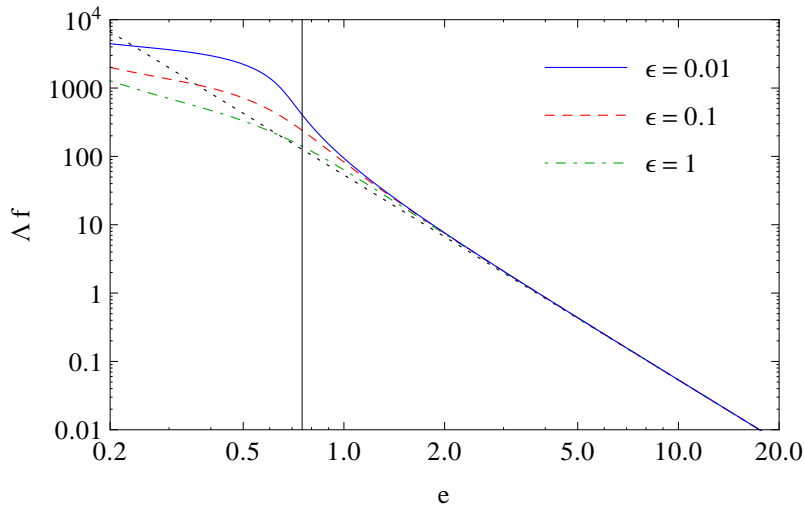}
	\caption{Value of $\Lambda$ given by equation~(\ref{eq:Lambda}) as a function of the Skyrme parameter $e$, for different values of $\epsilon = v^2 / f^2$. The straight dotted line corresponds to the ungauged case. The vertical bar at $e \cong 0.75$ is the value below which the integral diverges in the absence of the electroweak symmetry breaking term~(\ref{eq:epsilonterm}).}
	\label{fig:Lambda}
\end{figure}

The Lorentz translations and boosts lead to the quantisation of the skyrmion momentum and are irrelevant to our discussion. Similarly, the global transformation along the hypercharge generator has a trivial action on the classical skyrmion solution and thus does not play any physical role. On the contrary, spatial rotations and $SU(2)_W$ transformations are related respectively to the spin and isospin of the skyrmion. As mentioned above, due to the approximate spherical symmetry of the solution, these two transformation happen to be related, and we will see that the physical states share the same total spin and isospin quantum numbers. We make the following rotating skyrmion ansatz~\cite{Adkins:1983ya}
\begin{equation}
	\Phi(x,t) = R(t) \Phi(x) R(t)^\dag,
	\hspace{1cm}
	W_i(x,t) = R(t) W_i(x) R(t)^\dag,
\end{equation}
where the time-independent configurations $\Phi(x)$ and $W_i(x)$ are given by eq.~(\ref{eq:hedgehog}) and~(\ref{eq:Skyrme-Wu-Yang}) respectively, and
\begin{equation}
	R(t) = \exp\left[ i \theta_1(t) Q^3 \right] \exp\left[ i \theta_2(t) Q^2 \right] \exp\left[ i \theta_3(t) Q^3 \right]
\end{equation}
is the rotation matrix expressed in terms of the Euler angles $\theta_i(t)$, which are collective coordinates describing the skyrmion rotational degrees of freedom. The Lagrangian is then
\begin{equation}
	\Lagr = -E_0 + \Lambda \tr\left( \partial_0 R \, \partial_0 R^\dag \right)
		= -E_0 + \frac{1}{2} \Lambda \left( \dot\theta_i^2 + 2 \dot\theta_1 \dot\theta_3 \cos\theta_2 \right),
\end{equation}
where $E_0$ corresponds to the energy of the static skyrmion solution, given in figure~\ref{fig:M0}, and $\Lambda$ is a moment of inertia, given in the presence of just the Skyrme term by
\begin{equation}
	\Lambda = \frac{8 \pi}{3} \frac{1}{f e^3} \int\limits_0^\infty \d r \left[ r^2 \sin^2 F
		\left( \left( F' \right)^2 + 1 + (1+a)^2 \frac{\sin^2 F}{r^2} \right) + 2 \frac{e^2}{g^2} a^2 \right].
	\label{eq:Lambda}
\end{equation}
The numerical value of $\Lambda$ is displayed in figure~\ref{fig:Lambda}. Note that the dependence on $\epsilon$ comes at the lowest order only  implicitly through the functions $F$ and $a$. In the limit of large $e$, $\Lambda$ tends to $53.4 / \left( f e^3 \right)$. Upon canonical quantisation of the collective coordinates $\theta_i$, the Hamiltonian operator becomes
\begin{equation}
	H = E_0 + \frac{1}{2 \Lambda} \left[ \frac{1}{\sin^2\theta_2} \left( \frac{\partial^2}{\partial\theta_1^2}
		+ \frac{\partial^2}{\partial\theta_3^2} - 2 \cos\theta_2 \frac{\partial^2}{\partial\theta_1 \partial\theta_3} \right)
		+ \frac{1}{\sin\theta_2} \frac{\partial}{\partial\theta_2} \sin\theta_2 \frac{\partial}{\partial\theta_2} \right].
	\label{eq:Hamiltonian}
\end{equation}
The differential operator in $H$ is the Laplacian on a three-sphere, and its eigenfunctions are the Wigner $D$-functions $D^j_{m m'}(\theta_i)$ with eigenvalues
\begin{equation}
	E^j_{m m'} = E_0 + \frac{j (j+1)}{2 \Lambda}
	\label{eq:spectrum}
\end{equation}

The spin and isospin of the physical eigenstates are then defined as the conserved charge obtained from the Noether currents associated with spatial rotations and $SU(2)_W$ transformations respectively. Computing them, one obtains for the isospin $I$ and spin $J$
\begin{equation}
	I_k = i \Lambda \tr\left( R \, \partial_0 R^\dag Q^k \right),
	\hspace{1cm}
	J_k = -i \Lambda \tr\left( R^\dag \partial_0 R \, Q^k \right).
	\label{eq:spinisospin}
\end{equation}
The two operators are related to each other via a rotation as
\begin{equation}
	I_k = - \Omega_{kl} J_l,
	\hspace{1cm}
	\textrm{where~~}
	\Omega_{kl} = \frac{1}{2} \tr \left( Q^k R \, Q^l R^\dag \right) \in SO(3).
\end{equation}
This relation is particularly important, since it implies that $I^2 = J^2$, so that physical skyrmion states have the same total spin and isospin quantum numbers. The operator $I^2 = J^2$ turns out to be exactly the differential operator in~(\ref{eq:Hamiltonian}), with eigenvalues $j (j+1)$. The quantum numbers $m$ and $m'$ have  to be identified subsequently with the eigenvalues of the operators $I_3$ and $J_3$ respectively. $j$ takes either integer or half-integer values, depending if the skyrmion is a boson or a fermion, while $m$ and $m'$ run from $-j$ to $+j$ in integer steps. Note that, as mentioned before, the total spin-isospin equivalence is actually broken by the Higgs vacuum expectation value.

The bosonic or fermionic nature of skyrmions can depend on different factors. When the fourth homotopy group $\pi_4$ of the coset space is non-trivial and contains a $\mathbb Z_2$ subgroup, the soliton can be quantised either as a boson or as a fermion, depending on whether the two classes of mappings from space-time to the target space are associated with a relative minus sign or not~\cite{Finkelstein:1968hy}. However, for $SU(3)$ as for all relevant choices of cosets leading to a pseudo-Goldstone Higgs boson (see table~\ref{tab:homotopy}), $\pi_4(G/H)$ is trivial. This means that the spin statistics of the skyrmion is unambiguously fixed by the coefficient of the Wess-Zumino-Witten term: if $N_c$ is even, the skyrmion is a boson; if $N_c$ is odd, it is a fermion~\cite{Witten:1983tx}. This fact can be seen by considering an adiabatic 2$\pi$ rotation of the skyrmion in space, which yields a phase $i \pi N_c$ in the action through the Wess-Zumino-Witten term.

\begin{figure}
	\centering
	\includegraphics[width=0.7\linewidth]{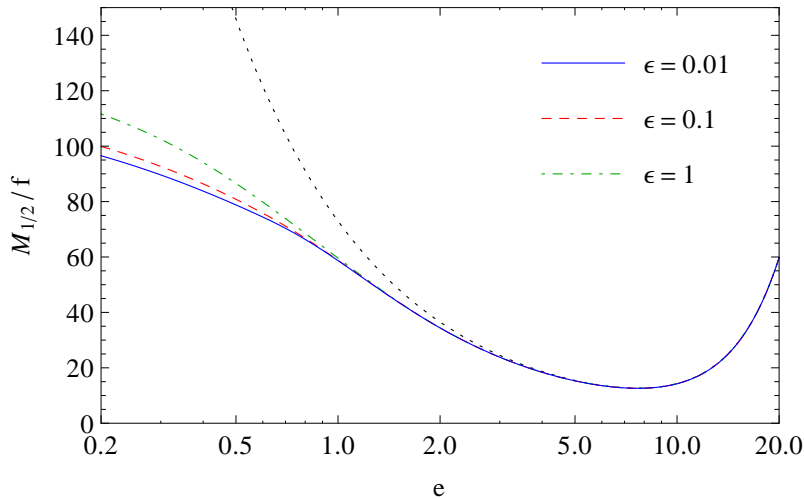}
	\caption{Mass of the lightest skyrmion state if it is a fermion as a function of the Skyrme parameter $e$, for different values of $\epsilon = v^2 / f^2$. The grey dotted line is the corresponding mass in the absence of gauge fields.}
	\label{fig:mass-fermion}
\end{figure}

If the skyrmion is a boson, the mass of the lowest state is given by $E_0$, displayed on figure~\ref{fig:M0} for the simplest case where only the Skyrme term is taken into account. On the contrary, if it is a fermion, its mass becomes
\begin{equation}
	M_\frac{1}{2} = E_0 + \frac{3}{8 \Lambda},
\end{equation}
and is shown on figure~\ref{fig:mass-fermion}. The different scalings of $E_0$ and $\Lambda^{-1}$ relative to the Skyrme parameter $e$ imply that a minimum occurs around $e_{min} \approx 7.7$, with $M_{\frac{1}{2},min} \approx 12.7~f$. The two lowest energy skyrmion states are then part of an isospin doublet, like the proton and the neutron in the Standard Model. Although the skyrmion mass is strongly dependent on the higher order terms in the sigma-model, its quantum numbers do not depend on the precise form of these terms, since the results rely only on the assumption that the solution is approximately spherically symmetric, which is always true when the skyrmion lives along the gauged $SU(2)_W$ subgroup of the global symmetry group.

\subsection{Electric charge of the skyrmion}

The hypercharge of the skyrmion can similarly be extracted from the Noether current associated with $U(1)_Y$ transformations. There is neither a contribution from the local Lagrangian given by the kinetic term~(\ref{eq:sigmamodel}) and the Skyrme term~(\ref{eq:Skyrmeterm}), nor from higher order operators, since the hypercharge generator commutes with the generators $T^a$ used to construct the hedgehog configuration. A non-zero contribution comes nevertheless from the non-local part of the Wess-Zumino-Witten term~(\ref{eq:WZWterm}), as discussed in the appendix. Regarding the skyrmion configuration living in the upper-left block of the matrix $\Phi$, the hypercharge generator appears as a multiple of the identity matrix, and therefore the contribution from the Wess-Zumino-Witten term is in the form of eq.~(\ref{eq:U1charge}), i.e. proportional to the number of colors $N_c$ of the underlying theory. In our toy model, the generator $Y$ given in~(\ref{eq:U1gaugegenerator}) comes with a factor $\frac{1}{6}$, so that the hypercharge of the skyrmion is fixed to be
\begin{equation}
	Y_\textrm{skyrmion} = \frac{N_c}{6}.
\end{equation}
The electric charge of the lightest skyrmion state, given by $Y + Q^3$, is then
\begin{equation}
	q = \left\{ \begin{array}{ll}
		\frac{1}{6} \, N_c & \textrm{if } N_c \textrm{ is even (boson),} \\
		& \\
		\frac{1}{6} \left( N_c \pm 3 \right) & \textrm{if } N_c \textrm { is odd (fermion).}
	\end{array} \right.
	\label{eq:electriccharge}
\end{equation}
There are subsequently only two cases in which the skyrmion can be electrically neutral: $N_c = 0$, for which the underlying theory is not strongly-coupled, and $N_c = 3$, as in QCD.

The skyrmion electric charge may also be understood in terms of the technifermions of the strongly-coupled theory. The form of the generator of electromagnetism~(\ref{eq:electricchargegenerator}) indicates that the technifermions have electric charges $\frac{2}{3}$ and $-\frac{1}{3}$. The skyrmion appears then as a bound state of $N_c$ technifermions. Either left-handed, right-handed, or both left- and right-handed technifermions  form doublets of $SU(2)_W$, and the lightest skyrmion state is then made of $N_c / 2$ ``up-type'' fermions and $N_c / 2$ ``down-type'' ones if $N_c$ is even, yielding the overall electric charge $q = N_c / 6$. If $N_c$ is odd, the two lightest states are made respectively of $\left( N_c \pm 1 \right) / 2$ ``up-type'' and $\left( N_c \mp 1 \right) / 2$ ``down-type'' technifermions, hence the result obtained in eq.~(\ref{eq:electriccharge}).

\subsection{A next-to-minimal model}

There is of course a simple possibility to relax the strong constraint~(\ref{eq:electriccharge}) on the electric charge by enlarging the model to a global $SU(N)_L \times SU(N)_R$ symmetry with $N \geq 4$. Keeping the $SU(2)_W$ subgroup to live in the upper-left corner, we can now take the hypercharge generator to be of the more general form
\begin{equation}
	Y = \left( \begin{array}{cccc}
		\mu &&& \\ & \mu && \\ && \mu-\frac{1}{2} & \\ &&& \ddots
	\end{array} \right),
\end{equation}
where factors denoted by the dots have to be chosen so that $Y$ remains traceless. This is the most general choice that produces a doublet with the correct hypercharge to play the role of a Higgs boson. A similar form can also be obtained within the simplest $N = 3$ model if one allows the hypercharge generator not to be traceless. This happens quite generically when the $U(1)_Y$ gauge group contains a part coming from the unbroken $U(1)_V$ symmetry. In both cases, all the analysis made above remains valid, up to the skyrmion hypercharge, which gets replaced by
\begin{equation}
	Y_\textrm{skyrmion} = \mu \, N_c.
\end{equation}
In terms of the electric charge, this means
\begin{equation}
	q = \left\{ \begin{array}{ll}
		\mu \, N_c & \textrm{if } N_c \textrm{ is even (boson),} \\
		& \\
		\mu \, N_c \pm \frac{1}{2} & \textrm{if } N_c \textrm { is odd (fermion).}
	\end{array} \right.
\end{equation}
If $N_c$ is even, the hypercharge of the skyrmion --- or equivalently its electric charge --- can always be made zero by choosing $\mu = 0$. The generator of electromagnetism is then
\begin{equation}
	Q = \left( \begin{array}{cccc}
		\frac{1}{2} &&& \\ & -\frac{1}{2} && \\ && -\frac{1}{2} & \\ &&& \ddots
	\end{array} \right),
\end{equation}
which reflects the electric charge of the technifermions. If $N_c$ is odd, a neutral skyrmion is achieved by taking $\mu = \pm 1 / \left( 2 N_c \right)$. There must thus be a weak doublet with charges $\left( 1 + N_c \right) / \left( 2 N_c \right)$ and  $\left( 1 - N_c \right) / \left( 2 N_c \right)$ (or opposite) among the technifermions of the underlying theory. In both cases, we see that the electric charges of the strongly-interacting technifermions have to be fractional in order to account for a neutral state among the lightest skyrmions.


\section{Other realisations of a composite Higgs}

There are many possible symmetry breaking patterns which enter the construction of a composite Higgs model. Most of them are presented in table~\ref{tab:homotopy} and will be now discussed separately. The presence of skyrmions depends on the third homotopy group of the vacuum manifold $G/H$. Note that in all examples presented here, the fourth homotopy group of the coset space is trivial, implying that the spin statistics of the skyrmion is entirely determined by the coefficient of the Wess-Zumino-Witten term, as for the model discussed above.

\subsection{SU(N)$\bm\times$SU(N)/SU(N)}

This is the symmetry breaking pattern of the chiral theory of mesons in QCD, which was used to construct the original technicolor models~\cite{Weinberg:1975gm, Susskind:1978ms}, as well as the ``minimal moose'' class of little Higgs models~\cite{ArkaniHamed:2002qx}. In the latter, many copies of the global symmetry need to be implemented, and there might also be many independent skyrmions, living in different sectors of the model. Their characteristic features follow directly from the toy model presented in section~\ref{sec:toymodel}.
\begin{table}
	\centering
	\renewcommand{\arraystretch}{1.2}
	\setlength{\tabcolsep}{20pt}
	\begin{tabular}{lcc}
		\hline
		\hline
		$G / H$ & $\pi_3 (G/H)$ & $\pi_4 (G/H)$ \\
		\hline
		$SU(N) \times SU(N) / SU(N)$ & $\mathbb{Z}$ & 0 \\
		$SO(N) \times SO(N) / SO(N)$ & $\mathbb{Z}$ & 0 \\
		$SU(N) / SO(N) ~~~~ N \geq 4$ & ~$\mathbb{Z}_2$ & 0 \\
		$SU(N) / SU(N-1)$ & 0 & 0 \\
		$SO(N) / SO(N-1)$ & 0 & 0 \\
		$SU(2N) / Sp(2N)$ & 0 & 0 \\
		\hline
		\hline
	\end{tabular}
	\caption{Third and fourth group of homotopy of different coset spaces $G/H$~\cite{Bryan:1993hz}.}
	\label{tab:homotopy}
\end{table}

\subsection{SO(N)$\bm\times$SO(N)/SO(N)}

The $SO(N) \times SO(N)/SO(N)$ symmetry breaking pattern is similar to the precedent, with the field $\Phi(x)$ being represented by a real-valued matrix in $SO(N)$. The representation of the $SU(2)_W$ gauge group has to be real and the Higgs field is not represented by a complex doublet, but rather by a real quadruplet. Little Higgs models based on this symmetry breaking pattern have the advantage of preserving a $SO(4)$ symmetry in the Goldstone sector, which acts as a custodial symmetry for the Higgs boson~\cite{Chang:2003un, Schmaltz:2010ac}. Note that the simplest real representation of $SU(2)$ based on the generators of $SO(3)$ does not allow the existence of such a quadruplet. We can take without loss of generality the gauge generators to be in the four-dimensional representation
\begin{equation*}
	Q^1 = \frac{1}{2} \left( \begin{array}{ccccc}
		&&& i & \\ && i && \\ & -i &&& \\ -i &&&& \\ &&&& \ddots
	\end{array} \right),
	\hspace{1cm}
	Q^2 = \frac{1}{2} \left( \begin{array}{ccccc}
		& -i &&& \\ i &&&& \\ &&& -i & \\ && i && \\ &&&& \ddots
	\end{array} \right),
\end{equation*}
\begin{equation}
	Q^3 = \frac{1}{2} \left( \begin{array}{ccccc}
		&& i && \\ &&& -i & \\ -i &&&& \\ & i &&& \\ &&&& \ddots
	\end{array} \right).
	\label{eq:SONgenerators}
\end{equation}
The construction of the lightest skyrmion follows similar principles as in the previous section, with a hedgehog ansatz living within the $SU(2)_W$ subgroup, i.e. choosing $T^a = Q^a$. The generators above obey the normalisation $\tr\left( Q^a Q^b \right) = \delta_{ab}$, which is twice the value required in~(\ref{eq:su2algebra}) to construct a skyrmion of unit winding number, but on the other hand the topological index in $SO(N)$ is defined as half the winding number integral~(\ref{eq:windingnumber}), so that the skyrmion can indeed be built out of gauge generators as in the toy model. A consequence of this peculiar normalisation is nevertheless that the energy of a static configuration is is multiplied by two with respect to the equivalent one in $SU(N)$. This modification doubles the mass of the skyrmion but is irrelevant to the computation of its quantum numbers.

The fundamental difference with respect to the toy model of section~\ref{sec:toymodel} is that the Wess-Zumino-Witten term vanishes when the field $\Phi(x)$ takes its values in $SO(N)$ as a consequence of the fact that the underlying theory is free of anomalies. Since there is no term in the action contributing a negative sign upon spatial rotation of the skyrmion, it must necessarily be quantised as a boson. Its hypercharge also vanishes independently of the implementation of the hypercharge generator and the skyrmion is automatically neutral, i.e.
\begin{equation}
	q = 0.
\end{equation}

\subsection{SU(N)/SO(N)}

The symmetry breaking pattern $SU(N) \to SO(N)$ is realised with $N=5$ in the ``littlest Higgs'' model~\cite{ArkaniHamed:2002qy}. The field $\Phi$ has to be taken in the two-indices symmetric representation of $SU(N)$, i.e. transforming as
\begin{equation}
	\Phi \to U \Phi U^\T,
	\hspace{1cm}
	U \in SU(5),
	\label{eq:Phi2indexsym}
\end{equation}
and a vacuum expectation value proportional to the identity breaks the global symmetry down to $SO(N)$. In this representation the definition of the covariant derivative must also be changed to
\begin{equation}
	D_\mu \Phi = \partial_\mu \Phi - i g A_\mu \left( Q \, \Phi + \Phi \, Q^\T \right),
\end{equation}
where $A_\mu$ denotes generically any gauge field. Such a symmetry breaking pattern can be achieved by considering technifermions transforming in the fundamental representation of $SU(N)$. The mechanism guaranteeing the quantum stability of the skyrmion in $SU(N)/SO(N)$ cosets is also completely different from the previous two cases, since there is no equivalent of a topological charge here~\cite{Auzzi:2006ns, Bolognesi:2007ut, Auzzi:2008hu, Bolognesi:2009vm}.

The generators of the weak gauge group can be taken as in~(\ref{eq:SONgenerators}) as a subgroup of $SO(4)$, but there is a unique implementation of the hypercharge --- up to global transformations --- yielding a quadruplet of $SO(4)$ which can be identified with the Higgs, namely
\begin{equation}
	Y = \frac{1}{2} \left( \begin{array}{ccccc}
		&& i && \\
		&&& i & \\
		-i &&&& \\
		& -i &&& \\
		&&&& \ddots 
	\end{array} \right).
\end{equation}
The Higgs appears then in the Goldstone field as
\begin{equation}
	\Pi = \left( \begin{array}{ccrc}
		&& h & \\
		&& -i \, h & \\
		h^\dag & i \, h^\dag && \\
		&&& \ddots
	\end{array} \right),
\end{equation}
where $h$ is the complex doublet defined in eq.~(\ref{eq:Higgsdoublet}).

The major difference with respect to our $SU(3)$-based example hides in the way the classical skyrmion solution is constructed. The embedding of the $SU(2)_W$ gauge generators in the unbroken $SO(N)$ subgroup of $SU(N)$ implies already that the skyrmion cannot be aligned with the gauge subgroup: due to the normalisation $\tr\left( Q^a Q^b \right) = \delta^{ab}$, the topological index of a hedgehog configuration built out of the generators $Q^a$ --- now measured again in $SU(N)$ by the winding number integral~(\ref{eq:windingnumber}) --- is always a multiple of two, and the configuration is therefore topologically trivial due to the $\mathbb Z_2$ homotopy structure of the vacuum manifold. A skyrmion of unit winding number is actually obtained~\cite{Gillioz:2010mr,Gillioz:2011tr} by the Cartan embedding of a $SU(N)$ hedgehog $\Phi_h$, as
\begin{equation}
	\Phi = \Phi_h \Phi_h^\T,
	\hspace{1cm}
	\textrm{where~~}
	\Phi_h = \exp\left[ 2 i \, F(r) \, \hat x_i \, T^i \right],
	\label{eq:Cartanembedding}
\end{equation}
with a special choice of generators allowing to preserve the spherical symmetry:
\begin{equation}
	T^a = \frac{1}{4} \left( \begin{array}{ccc}
		\sigma^a & i \sigma^a & \\
		-i \sigma^a & \sigma^a & \\
		&& \ddots
	\end{array} \right).
	\label{eq:SUNSONgenerators}
\end{equation}
Note that these generators have the property that they commute with their transpose, i.e.~$\left[ T^a, (T^b)^\T \right] = 0$, and are related to the $SU(2)_W$ gauge generators through
\begin{equation}
	Q^a = T^a - \left( T^a \right)^\T.
\end{equation}
The skyrmion field $\Phi$ constructed using the Cartan embedding~(\ref{eq:Cartanembedding}) lives actually in the broken subgroup of $SU(N)$. In the presence of the standard electroweak gauge field, the skyrmion configuration minimising the energy is obtained using the Skyrme-Wu-Yang ansatz~(\ref{eq:Skyrme-Wu-Yang}) for the gauge field, although the misalignment of the generators implies that the energy functional takes a form different from our toy model~(\ref{eq:energyfunctional}). In the littlest Higgs model with T-parity~\cite{Cheng:2004yc}, the skyrmion is however obtained in a slightly different way, due to the presence of an extra $SU(2)$ gauge group living in the broken part of the global $SU(N)$ symmetry. This gauge group is described by generators orthogonal to the $Q^a$, namely
\begin{equation}
	\bar{Q}^a = T^a + \left( T^a \right)^\T.
\end{equation}
The new gauge field $\bar{W}_\mu^a$ explicitly breaks the vacuum symmetry and obtains a mass proportional to the symmetry breaking scale $f$. The lightest skyrmion configuration is then a spherically symmetric configuration in which the electroweak gauge field vanishes, while an ansatz symmetric in the spatial and group indices is used for the new field~\cite{Gillioz:2010mr},
\begin{equation}
	\bar{W}_i^a = \frac{a(r)}{2 g r} \, \hat x_i \, \hat x_a,
	\hspace{1cm}
	\bar{W}_0^a = 0.
\end{equation}
However, the presence or absence of this extra spontaneously broken gauge field does not modify the way the skyrmion is embedded into $SU(N)$, so that the charge under the electroweak gauge group can be derived independently of the details of the model considered. The quantisation procedure follows a path similar to the discussion above. The relevant zero-modes in this case are spatial rotations only, which do not coincide with isospin transformations. The Hamiltonian and energy levels of the theory appear as in eq.~(\ref{eq:Hamiltonian}) and~(\ref{eq:spectrum}) respectively, with the slight difference that both the energy $E_0$ and kinetic momentum $\Lambda$ differ from the purely $SU(N)$ skyrmion. While $j$ and $m$ still correspond to the spin quantum numbers, taking (half-)integer values when the skyrmion is a boson (fermion), the isospin gets actually multiplied by a factor of two compared to eq.~(\ref{eq:spinisospin}), so that the relation between spin and isospin becomes
\begin{equation}
	I_k = - 2 \Omega_{kl} J_l,
	\hspace{1cm}
	\textrm{hence~~}
	I^2 = 4 J^2.
\end{equation}
The third component of isospin of the skyrmion state is then given by $2 m'$, where $m'$ takes integer steps from $-j$ to $+j$. The lightest skyrmion state has thus zero isospin as before if it is a boson. If it is quantised as a fermion, the skyrmion has however an isospin number $I_3 = \pm 1$ and its electric charge is equal to its hypercharge plus or minus one unit. The hypercharge itself is uniquely fixed: it is computed similarly as above using the Noether currents associated with the $U(1)_Y$ transformation. Due to the commutation relations $\left[ Y, T^a \right] = \left[ Y, (T^a)^\T \right] = 0$, the skyrmion is again invariant under hypercharge transformations, and none of the local terms in the Lagrangian contribute to the Noether current. Still, the five-dimensional part of the Wess-Zumino-Witten term gives the usual non-zero contribution, fixing
\begin{equation}
	Y_\textrm{skyrmion} =  \pm N_c.
	\label{eq:SUNSONhypercharge}
\end{equation}
The plus or minus sign here refers to the fact that there are two distinct possibilities to construct the skyrmion field, namely using either the generators $T^a$ as above or the transpose thereof, $(T^a)^\T$, or equivalently choosing the $SU(N)$ hedgehog configuration $\Phi_h$ in eq.~(\ref{eq:Cartanembedding}) to have plus or minus one unit of winding number. Note that apart from their hypercharge, both field configurations are absolutely identical at the classical level. The skyrmion charge associated with the solution of hypercharge $+N_c$ is then
\begin{equation}
	q = \left\{ \begin{array}{ll}
		N_c & \textrm{if } N_c \textrm{ is even (boson),} \\
		& \\
		N_c \pm 1 & \textrm{if } N_c \textrm { is odd (fermion).}
	\end{array} \right.
	\label{eq:SUNSONcharge}
\end{equation}
As a direct consequence, there are only two possibilities to have a neutral skyrmion: the first consists in taking $N_c = 0$, i.e. considering a weakly-coupled UV completion and quantising the skyrmion as a boson, and the second to $N_c = 1$, which does not correspond to a satisfactory strongly-coupled UV completion either. Note that a weakly-coupled UV completion exists for the littlest Higgs~\cite{Csaki:2008se}, but the quantum nature of the skyrmion with both $N_c = 0,1$ is still not understood at all. On the other hand, any strongly-coupled UV completion with $N_c \geq 2$ possesses a charged skyrmion in its spectrum. The integer value of the charge can be found surprising at first. However, it can be understood in terms of the technifermions of a strongly-coupled underlying theory. To see this, it is convenient to rotate the vacuum expectation value $\langle\Phi\rangle = \identity$ into a basis where the generators $Q^3$ and $Y$ are diagonal. This can be done as in the original little Higgs paper~\cite{ArkaniHamed:2002qy}, where we have
\begin{equation}
	\left. \begin{array}{lll}
		Q^3 & = & \textrm{diag}\left( \frac{1}{2}, -\frac{1}{2}, \frac{1}{2}, -\frac{1}{2}, 0, \ldots \right) ~ \\
		\\
		Y & = & \textrm{diag}\left( -\frac{1}{2}, -\frac{1}{2}, \frac{1}{2}, \frac{1}{2}, 0, \ldots \right) ~
	\end{array}\right\}
	\Longrightarrow
	\hspace{0.5cm}
	Q_{em} = \textrm{diag}\left( 0, -1 , 1, 0, \ldots \right).
\end{equation}
We see that the technifermions all have necessarily integer electric charge in this model, which is consistent with the integer charge of the skyrmion. Note that in contrary to the previous cases, the skyrmion in the $SU(N)/SO(N)$ coset is not made of an antisymmetric bound state of $N_c$ technifermions, but appears as a bound state of $N_c \left( N_c + 1 \right) / 2$ of them~\cite{Auzzi:2006ns, Bolognesi:2007ut, Auzzi:2008hu, Bolognesi:2009vm}. A skyrmion of charge $N_c$ can for example be constructed with $N_c$ positively charged technifermions and $N_c \left( N_c - 1 \right) / 2$ neutral ones, although other combinations are possible.

\subsection{Cosets with trivial homotopy}

The remaining symmetry patterns presented in table~\ref{tab:homotopy} all have a trivial third homotopy group, and therefore do not contain skyrmions and are not concerned by our study. The model dubbed ``little Higgs from a simple group''~\cite{Kaplan:2003uc}, based on the coset $SU(N)/SU(N-1)$, is an example of a model free of skyrmions. The field $\Phi$ is taken there to be a $N$-component vector of unit length, transforming under $SU(N)$ and spontaneously breaking the symmetry by choosing any definite value in the vacuum. The same concept applies when the field $\Phi$ is real-valued, hence providing a $SO(N)/SO(N-1)$ vacuum manifold, as in the minimal composite Higgs model~\cite{Agashe:2004rs} and its extension~\cite{Gripaios:2009pe}, with $N=5$ and $6$ respectively. The last coset mentioned in the table, namely $SU(2N)/Sp(2N)$, is realised in the ``little Higgs from an antisymmetric condensate''~\cite{Low:2002ws}.

Note that the list in table~\ref{tab:homotopy} is not exhaustive. Other cosets may be considered, in particular when $H$ is not a simple group (see for example~\cite{Rychkov:2011br} for a more complete list of possible symmetry breaking patterns). Some of these cosets might also admit skyrmions.


\section{Summary and consequences}

The results of the previous sections can be summarised into three main points:

\begin{enumerate}[(i)]

\item
Skyrmions can generally use gauge fields to reduce their mass. The lightest field configuration of unit winding number is obtained at the classical level by combining the Goldstone fields with a $SU(2)$ gauge field, independently of the coefficients of the higher-order terms in the sigma-model. This configuration also preserves generically the spherical symmetry of the ungauged skyrmion. Notice that this statement is universal, and that it does not matter if the $SU(2)$ subgroup is fundamental to the theory, or if it arises as the unbroken remnant of a larger gauge group.

\item
In the real world, there is no such exact $SU(2)$ symmetry, since the electroweak gauge group is broken down to $U(1)_{em}$. However, at the scale 10--100~TeV where the skyrmion lives, the $SU(2)_W \times U(1)_Y$ gauge symmetry is approximately preserved, up to corrections of order $v^4 / f^4$, where $f$ is the TeV-order symmetry breaking scale. As a consequence, the skyrmion in composite Higgs models prefers to live in a subspace of the global symmetry where it can use this $SU(2)_W$ symmetry.

\item
The isospin of the skyrmion depends on its fermionic/bosonic nature, which is fixed via the Wess-Zumino-Witten term by the number of colors of the underlying strongly-coupled theory (if any). It is also the non-local part of this Wess-Zumino-Witten term which contributes to the hypercharge of the skyrmion, so that the electric charge of the lightest state(s) is completely fixed by the embedding of the hypercharge generator relatively to the $SU(2)_W$ subgroup.

\end{enumerate}
We have studied various cosets in this work, all presented in table~\ref{tab:homotopy}, and found that two of them contain in general electrically charged skyrmions. Unfortunately, these two are also the most used symmetry breaking patterns. In the various littlest Higgs models, where the vacuum manifold is $SU(5)/SO(5)$, the skyrmion takes an integer electric charge, necessarily different from zero for $N_c \geq 2$. For $SU(N) \times SU(N)/SO(N)$ cosets which mimic the chiral theory of QCD, neutral skyrmions can be obtained, provided certain rules are respected concerning the embedding of the hypercharge in the global symmetry group. In terms of degrees of freedom of the underlying theory, a neutral skyrmion requires that the technifermions have fractional electric charges. As a general result, weakly coupled UV completions may always admit neutral skyrmions, since non-zero charges are induced via the anomalous Wess-Zumino-Witten term present in strongly-coupled theories only; in that case however it is not clear what the skyrmion really represents at the quantum level of the theory. Also if the underlying theory is strongly-coupled but still anomaly free, as for the $SO(N) \times SO(N)/SO(N)$ cosets realised in models with a custodial symmetry, the skyrmion has to be quantised as a boson and is automatically neutral. Other models based on the remaining symmetry breaking patterns don't contain skyrmions at all.

An estimate of the relic density for a neutral skyrmion was made in ref.~\cite{Gillioz:2010mr}. The key point here is that skyrmion are always expected to be thermally produced in the early universe, while their pair-annihilation cross-section is naturally very low --- scaling with the inverse of the symmetry breaking scale $f$ --- which ensure relatively large relic densities. This relic density at present time can be of the order of the observed dark matter density, although much smaller or larger densities can also be obtained depending on the exact value of the skyrmion mass which cannot be computed without knowing the full underlying theory.

In the presence of an electrically charged skyrmion, the situation is very different. Due to electromagnetic interactions, the skyrmion can form bound states with ordinary matter. If the skyrmion charge is an integer, neutral bound states are formed and can be very difficult to detect. The strongest experimental bounds come from the concentration of such super-heavy ``atoms'' on earth, as studied for example in ref.~\cite{Hemmick:1989ns,Smith:1982qu}. Although the relevant properties of these skyrmions are simply their mass and electric charge, it is not straightforward to extract experimental bounds on these parameters, due to the fact that the whole early universe cosmology can be affected by the presence of new stable charged objects (see for example the effects of a stable particle of charge $-2$ in ref.~\cite{Khlopov:2007ic}). Note also that the skyrmions described in this work always come in pairs with an antiskyrmion of opposite charge: in the scenario where the topological index takes any integer value, the antiskyrmion is a field configuration with opposite winding number; for $SU(N)/SO(N)$ with $\pi_3 = \mathbb Z_2$, the partner of the skyrmion comes actually from a different embedding into the coset. In any case, if no asymmetry is generated, the relic density of skyrmions might be largely reduced compared to a naive estimate.

In contrast to little Higgs models, composite Higgs models arising from the holographic picture are usually built using cosets which do not admit skyrmions. This is however not a necessary condition, and composite Higgs models containing skyrmions could also be built. The crucial difference with respect to little Higgs models is that the technifermions have to share the same quantum number as the Standard Model quarks and leptons so that they can generate masses with the latter via kinetic mixing. Neutral skyrmions should subsequently be easier to obtain in composite Higgs models, and would appear as heavy objects bounded by strong interactions in a similar fashion as the proton in QCD.

Our analysis also contains a few caveats, since the description of non-perturbative objects such as skyrmions in quantum field theories is not well understood yet. The conclusions of this work might be relaxed under special conditions. First of all, we actually only understand the quantum nature of the skyrmion in the background of a strongly-coupled underlying theory. In the cases $N_c = 0$ or $N_c = 1$ discussed above, the role of the skyrmion in the spectrum of the quantum theory is not known: the skyrmion might actually be unstable despite its topological nature, or might  even not correspond to a particle state. The second unknown concerns the regime in which the level splitting between the various skyrmion states is very small. Quantum corrections to the skyrmion masses could possibly reverse the hierarchy of the states, making practically impossible any predictions in terms of the low-energy effective theory. Loop corrections to the skyrmion mass in the chiral theory of QCD happen indeed to be important~\cite{Meier:1996ng}, but one can expect them to act in an universal way on all states so that such a hierarchy-reversing phenomenon is rather unlikely to occur.

Note finally that for composite models based on the holographic principle, where an extra-dimensional description of the strongly-coupled underlying theory exists, the properties of the skyrmion might be better understood by directly computing them within the five-dimensional framework, in the light of recent developments in this field~\cite{Pomarol:2007kr}. This subject remains open for future analysis.


\subsection*{Acknowledgements}

I am thankful to Pedro Schwaller, Andreas von Manteuffel and Daniel Wyler for their collaboration at an early stage of this work and to Erich Weihs for useful comments on the manuscript. This work was supported by the Schweizer Nationalfonds and by the European Commission through the ``LHCPhenoNet'' Initial Training Network PITN-GA-2010-264564.

\clearpage
\appendix
\section*{Appendix: Wess-Zumino-Witten term and Noether currents in the presence of vector gauge fields}

The non-local five-dimensional part $\Gamma_5$ of the Wess-Zumino-Witten term~(\ref{eq:WZWterm}) is not gauge invariant. However, its variation under a vector gauge transformation
\begin{equation}
	\Phi \to U \Phi U^\dag,
	\hspace{1cm}
	U \in SU(N),
\end{equation}
can be written as a total derivative:
{\setlength\arraycolsep{1.4pt}
\begin{eqnarray}
	\delta\Gamma_5 & = & \frac{i N_c}{48 \pi^2}  \int_{\mathcal{M}_5} \d \left[
		\tr\left( U^\dag \d U \, \Phi \, \d\Phi^\dag \d\Phi \, \d\Phi^\dag \right)
		+ \frac{1}{2} \tr\left( U^\dag \d U \, \Phi \, \d\Phi^\dag U^\dag \d U \, \Phi \, \d\Phi^\dag \right)
		\right. \nonumber \\
	&& \hspace{2.3cm} \left.
		+ \tr\left( U^\dag \d U \, \Phi \, U^\dag \d U \, \Phi^\dag \d\Phi \, \d\Phi^\dag \right)
		+ \tr\left( U^\dag \d U \, \d U^\dag \d U \, \Phi \, \d\Phi^\dag \right)
		\right. \nonumber \\
	&& \hspace{2.3cm} \left.
		+ \tr\left( \d U^\dag \d U \, \Phi \, U^\dag \d U \, \d\Phi^\dag \right)
		+ \tr\left( U^\dag \d U \, \d U^\dag \d U \, \Phi \, U^\dag \d U \, \Phi^\dag \right)
		\right. \nonumber \\
	&& \hspace{2.3cm} \left.
		+ \tr\left( U^\dag \d U \, \Phi \, U^\dag \d U \, \Phi^\dag U^\dag \d U \, \Phi \, \d\Phi^\dag \right)
		\right. \nonumber \\
	&& \hspace{2.3cm} \left.
		+ \frac{1}{4}
			\tr\left( U^\dag \d U \, \Phi \, U^\dag \d U \, \Phi^\dag U^\dag \d U \, \Phi \, U^\dag \d U \, \Phi^\dag \right)
		- \left( \Phi \leftrightarrow \Phi^\dag \right)
		\right]
\end{eqnarray}}%
where we used the language of differential forms, i.e. $\d\Phi = \partial_\mu \Phi \, \d x^\mu$, $\d U = \partial_\mu U \, \d x^\mu$, and the antisymmetrisation of indices is implicitly understood. Using Stokes' theorem, and since the boundary of $\mathcal{M}_5$ is the usual Minkowski space $\mathcal{M}_4$, $\delta\Gamma_5$ can be expressed as a four-dimensional integral. A gauge invariant form of the Wess-Zumino-Witten can thus be obtained by adding a four dimensional action written in terms of the scalar field $\Phi$ and of the gauge field $A = A_\mu \, \d x^\mu$, transforming under the $SU(N)$ vector gauge transformation as
\begin{equation}
	A_\mu \to U A_\mu U^\dag + \frac{i}{g} U \partial_\mu U^\dag.
\end{equation} 
The strategy to determine this gauge counterterm is to write all possible 4-forms in terms of $\Phi$ and $A$, and to choose the coefficients of each of them so that the total variation exactly cancels $\delta\Gamma_5$. We find
{\setlength\arraycolsep{1.4pt}
\begin{eqnarray}
	\Gamma_4 & = & \frac{i N_c}{48 \pi^2} \int_{\mathcal{M}_4} \Big[ 
		i g \tr\left( A \, \Phi^\dag \d\Phi \, d\Phi^\dag \d\Phi \right)
		+ g^2 \tr\left( (\d A \, A + A \, \d A) \Phi^\dag \d\Phi \right)
		\Big. \nonumber \\
	&& \hspace{2cm} 
		+ g^2 \tr\left( \d A \, \d\Phi^\dag A \, \Phi \right)
		+ g^2 \tr\left( A \, \Phi^\dag A \Phi \, \d\Phi^\dag \d\Phi \right)
		 \nonumber \\
	&& \hspace{2cm} 
		+ \frac{1}{2} g^2 \tr\left( A \, \Phi^\dag \d\Phi \, A \Phi^\dag \d\Phi \right)
		+ i g^3 \tr\left( A^3 \Phi \, \d\Phi^\dag \right)
		 \nonumber \\
	&& \hspace{2cm} 
		+ i g^3 \tr\left( (\d A \, A + A \, \d A) \Phi \, A \, \Phi^\dag \right)
		+ i g^3 \tr\left( A \, \Phi \, A \, \Phi^\dag A \, \Phi \, \d\Phi^\dag \right)
		 \nonumber \\
	&& \hspace{2cm} 
		+ g^4 \tr\left( A^3 \Phi \, A \, \Phi^\dag \right)
		+ \frac{1}{4} g^4 \tr\left( A \, \Phi^\dag A \, \Phi \, A \, \Phi^\dag A \, \Phi \right)
		- \left( \Phi \leftrightarrow \Phi^\dag \right)
		 \nonumber \\
	&& \hspace{2cm} 
		+ c_1 \, g^2 \tr \d\left( A^2 \Phi^\dag \d\Phi \right)
		+ c_2 \, g^2 \tr \d\left( A^2 \Phi \, \d\Phi^\dag \right)
		 \nonumber \\
	&& \hspace{2cm} 
		+ c_3 \, g^2 \tr \d\left( \d A \, \Phi^\dag A \, \Phi \right)
		+ c_4 \, g^2 \tr \d\left( \d A \, \Phi \, A \, \Phi^\dag \right)
		 \\
	&& \hspace{2cm} 
		+ c_5 \, g^3 \tr \d\left( A^2 \Phi^\dag A \, \Phi \right)
		+ c_6 \, g^3 \tr \d\left( A^2 \Phi \, A \, \Phi^\dag \right)
		 \nonumber \\
	&& \hspace{2cm} 
		+ c_7 \, g^2 \tr \d\left( A \, \Phi^\dag A \, \d\Phi + A \, \Phi \, A \, \d\Phi^\dag \right)
		+ c_0 \, g^2 \tr\left( F \, \Phi \, F \, \Phi^\dag \right) \Big.\Big], \nonumber
\end{eqnarray}}%
where $F = \d A - i g A^2$. The eight coefficients $c_i$ can in principle be chosen freely, since the variation of the corresponding operators is zero (for the $c_0$ term) or vanish upon integration ($c_1$ to $c_7$ terms). The total Wess-Zumino-Witten action
\begin{equation}
	\Gamma_{WZW} = \Gamma_5 + \Gamma_4
\end{equation}
is then gauge-invariant.

The Noether currents associated with a global $SU(N)_V$ infinitesimal transformation defined as 
\begin{equation}
	\delta\Phi = i \left[ T, \Phi \right]
	\hspace{1cm}
	\textrm{and}
	\hspace{1cm}
	\delta A_\mu = i \left[ T, A_\mu \right],
\end{equation}
where $T$ is some $SU(N)$ generator, is then
{\setlength\arraycolsep{1.4pt}
\begin{eqnarray}
	J^\mu_{WZW} & = & \frac{N_c}{48 \pi^2} \epsilon^{\mu\nu\rho\sigma} \Big[
		\tr\left( T \, \Phi^\dag \partial_\nu \Phi \, \partial_\rho \Phi^\dag \partial_\sigma \Phi \right)
		+ i g \tr\left( T \, A_\nu \partial_\rho \Phi \, \partial_\sigma \Phi^\dag \right)
		\Big. \nonumber \\
	&& \hspace{2cm} 
		+ i g \tr\left( T \, \partial_\nu \Phi^\dag A_\rho \partial_\sigma \Phi \right)
		+ i g \tr\left( T \, \partial_\nu \Phi \, \partial_\rho \Phi^\dag A_\sigma \right)
		 \nonumber \\
	&& \hspace{2cm} 
		+ i g \tr\left( T \, \Phi \, A_\nu \Phi^\dag \partial_\rho \Phi \, \partial_\sigma \Phi^\dag \right)
		+ i g \tr\left( T \, \Phi \, \partial_\nu \Phi^\dag A_\rho \Phi \, \partial_\sigma \Phi^\dag \right)
		 \nonumber \\
	&& \hspace{2cm} 
		+ i g \tr\left( T \, \Phi \, \partial_\nu \Phi^\dag \partial_\rho \Phi \, A_\sigma \Phi^\dag \right)
		+ g^2 \tr\left( T \, A_\nu A_\rho \Phi \, \partial_\sigma \Phi^\dag \right)
		 \nonumber \\
	&& \hspace{2cm} 
		+ g^2 \tr\left( T \, A_\nu \Phi^\dag \partial_\rho \Phi \, A_\sigma \right)
		+ g^2 \tr\left( T \, \Phi \, \partial_\nu \Phi^\dag A_\rho A_\sigma \right)
		 \nonumber \\
	&& \hspace{2cm} 
		+ g^2 \tr\left( T \, A_\nu \Phi^\dag A_\rho \partial_\sigma \Phi \right)
		+ g^2 \tr\left( T \, A_\nu \partial_\rho \Phi \, A_\sigma \Phi^\dag \right)
		 \nonumber \\
	&& \hspace{2cm} 
		+ g^2 \tr\left( T \, \Phi \, \partial_\nu \Phi^\dag A_\rho \Phi \, A_\sigma \Phi^\dag \right)
		+ g^2 \tr\left( T \, \Phi \, A_\nu \Phi^\dag \partial_\rho \Phi \, A_\sigma \Phi^\dag \right)
		 \nonumber \\
	&& \hspace{2cm} 
		+ g^2 \tr\left( T \, \Phi \, A_\nu \Phi^\dag A_\rho \Phi \, \partial_\sigma \Phi^\dag \right)
		+ g^2 \tr\left( T \, A_\nu \Phi \, \partial_\rho A_\sigma \Phi^\dag \right)
		 \nonumber \\
	&& \hspace{2cm} 
		+ g^2 \tr\left( T \, \Phi \, A_\nu \Phi^\dag \partial_\rho A_\sigma \right)
		+ g^2 \tr\left( T \, \Phi \, A_\nu \partial_\rho A_\sigma \Phi^\dag \right)
		 \nonumber \\
	&& \hspace{2cm} 
		+ g^2 \tr\left( T \, \Phi \, \partial_\nu A_\rho A_\sigma \Phi^\dag \right)
		+ i g^3 \tr\left( T \, A_\nu A_\rho \Phi^\dag A_\sigma \Phi \right)
		 \nonumber \\
	&& \hspace{2cm} 
		+ i g^3 \tr\left( T \, A_\nu \Phi \, A_\rho \Phi^\dag A_\sigma \right)
		+ i g^3 \tr\left( T \, \Phi^\dag A_\nu \Phi \, A_\rho A_\sigma \right)
		 \nonumber \\
	&& \hspace{2cm} 
		+ i g^3 \tr\left( T \, \Phi^\dag A_\nu A_\rho A_\sigma \Phi \right)
		+ i g^3 \tr\left( T \, \Phi^\dag A_\nu \Phi \, A_\rho \Phi^\dag A_\sigma \Phi \right)
		 \nonumber \\
	&& \hspace{2cm}
		-  \left( \Phi \leftrightarrow \Phi^\dag \right)
		\Big.\Big].
	\label{eq:Noethercurrents}
\end{eqnarray}}%
The first term of the integrand comes from the five-dimensional action $\Gamma_5$, all the remaining ones from $\Gamma_4$.

Note that for a field configuration $\Phi_0$ which is not charged under the generator $T$, i.e. $\left[ T, \Phi_0 \right]$, the variation $\delta\Phi_0$ vanishes, and so do all the terms in the current coming from the local part of the Wess-Zumino-Witten action. However, the five-dimensional terms can still give a contribution. Similarly, taking $T = \identity$ the Noether current does not vanish although the variation $\delta\Phi_0$ is zero, and we have
\begin{equation}
	J^\mu_{WZW,\identity} = \frac{N_c}{24 \pi^2} \epsilon^{\mu\nu\rho\sigma} 
		\tr\left( \Phi^\dag \partial_\nu \Phi \, \partial_\rho \Phi^\dag \partial_\sigma \Phi \right).
\end{equation}
The associated conserved charge
\begin{equation}
	Q_{\identity} = \int \d^3x \, J^0_{WZW,\identity}
		= \frac{N_c}{24 \pi^2} \int \d^3x \, \epsilon_{ijk} \tr\left( \Phi^\dag \partial_i \Phi \, \partial_j \Phi^\dag \partial_k \Phi \right)
		= N_c \, \eta
	\label{eq:U1charge}
\end{equation}
is exactly $N_c$ times the winding number integral $\eta$ given in eq.~(\ref{eq:windingnumber}).


\bibliographystyle{utphys}
\bibliography{Bibliography}

\clearpage

\section*{Erratum}

There is a confusion in the original paper between the coefficient multiplying the Wess--Zumino--Witten term, Eq.~(\ref{eq:WZWterm}), and the number of colors of the underlying strongly--coupled field theory. The symbol $N_c$ is used for both, but only coincide if the technifermions transform in the fundamental representation of the gauge group. Moreover, the coefficient entering Eq.~(\ref{eq:WZWterm}), for which we shall keep the notation $N_c$, was claimed to take integer values from topological considerations. However, this depends actually on the form of the matrix $\Phi$. For example, when $\Phi$ transforms in the two--index symmetric representation of $SU(N)$, as in Eq.~(\ref{eq:Phi2indexsym}), $N_c$ can take half--integer values.

The correct relation between the coefficient $N_c$ and the true number of colors of the underlying theory, call it $n_c$, is dependent on the representation of the technifermions. If the latter are Dirac fermions transforming in the fundamental of a $SU(N_c)$ gauge group, one has, as mentioned above,
\begin{equation}
	N_c = n_c \quad (\textrm{fundamental}).
\end{equation}
If on the contrary they are Weyl fermions transforming in the adjoint representation, as in Refs.~\cite{Auzzi:2006ns, Bolognesi:2007ut, Auzzi:2008hu, Bolognesi:2009vm}, one finds
\begin{equation}
	N_c = \frac{n_c^2 - 1}{2} \quad (\textrm{adjoint}).
\end{equation}
Similar relations can be established for higher--dimensional representations.

The conclusions of the paper remain unchanged. Only the interpretation of the quantity denoted by $N_c$ is altered. In terms of the true number of colors $n_c$, Eq.~(\ref{eq:SUNSONcharge}) would become for example
\begin{equation}
	q = \left\{ \begin{array}{ll}
		\frac{n_c^2 - 1}{2} & \textrm{if } n_c^2 - 1 \textrm{ is even (boson),} \\
		& \\
		\frac{n_c^2 - 1}{2} \pm 1 & \textrm{if } n_c^2 - 1 \textrm { is odd (fermion),}
	\end{array} \right.
\end{equation}
which is anyway incompatible with $q = 0$ for $n_c \geq 2$. The possible half--integer value of the electric charge for the skyrmion is not in contradiction with the integer electric charge of the Goldstone bosons, since in this case massive fermions must be added to the effective Lagrangian description, as argued in Refs.~\cite{Auzzi:2006ns, Bolognesi:2007ut, Auzzi:2008hu, Bolognesi:2009vm}.

\end{document}